\documentclass[9pt, A4]{article}
\usepackage[left=2.5cm, right=2.5cm, top=1.785cm, bottom=2.0cm]{geometry}
\usepackage{amsmath}
\usepackage[utf8]{inputenc}
\usepackage{graphicx} 
\usepackage[usenames,dvipsnames]{xcolor}
\usepackage{caption}

\title{The role of temperature in the rigidity-controlled fracture of elastic networks}
\author{Justin Tauber, Aim\'ee R. Kok, Jasper van der Gucht,$^{\ast}$ and Simone Dussi}
\date{}

\begin{document}

\maketitle

\begin{abstract}
    We study the influence of thermal fluctuations on the fracture of elastic networks, via simulations of the uniaxial extension of central-force spring networks with varying rigidity, i.e. connectivity. Studying their failure response, both at the macroscopic and microscopic level, we find that an increase in temperature corresponds to a more homogeneous stress (re)distribution and induces thermally activated failure of springs. As a consequence, the material strength decreases upon increasing temperature, the damage is spread over larger lengthscales and a more ductile fracture process is observed. These effects are modulated by network rigidity and can therefore be tuned via the network connectivity and the rupture threshold of the springs. Knowledge of the interplay between temperature and rigidity improves our understanding of the fracture of elastic networks, such as (biological) polymer networks, and can help to refine design principles for tough soft materials.
\end{abstract}

\section{\label{sec:intro}Introduction}
\begin{figure}[!ht]
\centering
\includegraphics{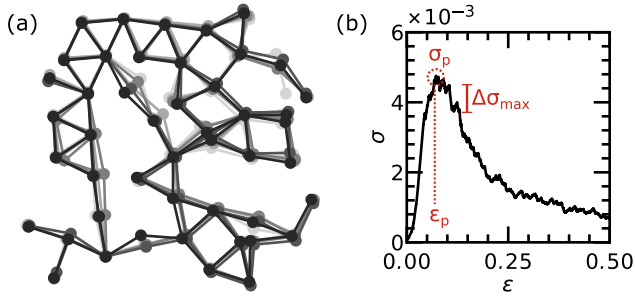}
\caption{\label{fig:introfig} Fracture of thermal spring networks. (a) Portion of spring network under 1.5\% extensional strain. Several snapshots (corresponding to different shades of gray) are overlaid to indicate the effect of thermal motion. (b) Example of stress-strain response of a diluted spring network ($p=0.65$, $\lambda=0.03$, $T^{*}= 10^{-4}$, $L=128$). We highlight the peak-stress $\sigma_p$, the corresponding peak strain $\varepsilon_p$ and the maximum drop in stress $\Delta\sigma_{\text{max}}$.}
\end{figure}

Many soft materials, such as (biological) hydrogels, elastomers and colloidal gels have an underlying network structure~\cite{Jansen2018,Kouwer2013,Creton2017,Trappe2001}. The structure of these materials can be reduced to a collection of elements or springs interconnected by cross-links or nodes, as schematically shown in Fig.~\ref{fig:introfig}(a), to study the contribution of network structure to the properties of these materials \cite{Alava2006,Picu2011,Broedersz2014}. Recent computational studies, complemented with experiments on architectured elastic networks, have shown that under athermal and quasistatic conditions the elastic response and failure behaviour of an elastic network is controlled by the rigidity of the network and the strength of the individual elements~\cite{Driscoll2016,Zhang2016,Berthier2019,Dussi2020}. The network architecture, in particular its connectivity, determines the network rigidity. Commonly, the average connectivity of a random network is described using the average number of bonds per crosslink. Central-force spring networks, where the elements only resist stretching, are rigid above a connectivity $2d$, with $d$ the dimensionality of the network, called the isostatic point~\cite{Maxwell1864}. Below the isostatic point, central-force spring networks are mechanically floppy. However, simulations have shown that even floppy or sub-isostatic networks can be rigidified by an external deformation~\cite{Wyart2008,Shivers2019a,Dussi2020} or by the presence of additional interactions such as a bending rigidity~\cite{Broedersz2011,Sharma2016,Feng2016,Bouzid2018}. These studies suggest that if the static network structure and the element strength of a soft material are known, the material response can be predicted. Experiments indicate that for stiff fiber networks, such as collagen, this structure-property relation exists for both the elastic response~\cite{Sharma2016,Jansen2018} and the failure behaviour~\cite{Burla2020}.

Yet, we know that many soft materials are sensitive to thermal fluctuations, i.e. they do not satisfy the athermal limit assumed in the studies discussed above. This is most prevalent at the element level, for example, in the entropic elasticity of a polymer chain. In fact, a description of the mechanical response of polymer networks often only includes the properties of the elastic elements and neglects contributions of the network structure ~\cite{James1943,Arruda1993}. However, simulations on central-force spring networks suggest that in the presence of thermal fluctuations the effect of temperature on the linear modulus is rigidity dependent. In particular, it is shown that thermal fluctuations can stabilize (sub-isostatic) central-force spring networks at the network level in a similar way to bending interactions~\cite{Dennison2013a,Zhang2016}. Also, when looking at failure, the average external force required to break a single element or bond is typically reduced in presence of thermal fluctuations.~\cite{Bell1978,Bullerjahn2014}. Experiments on polymer networks~\cite{Skrzeszewska2010} not only demonstrate the presence of thermally activated failure in network materials, but also imply that in networks the thermally activated failure is enhanced. These results suggest that, when thermal fluctuations are present, the elasticity and failure response are controlled by both network rigidity and thermal fluctuations. However, it remains unclear if, similar to the linear elastic regime, the impact of temperature couples to the rigidity of the network.

In this paper we explore to what extent network rigidity controls the influence of thermal fluctuations on the failure behaviour of an elastic material. To this end, we study the response of diluted central-force spring networks (see Fig.~\ref{fig:introfig}), similar to previous studies~\cite{Dennison2013a,Zhang2016,Driscoll2016,Dussi2020}. To introduce thermal fluctuations into these systems we perform Langevin Dynamics simulations, in which an implicit solvent keeps the temperature of the system constant. We find that the strength of the networks is dependent on temperature and that the effect of the thermal fluctuations is coupled to the rigidity of the network. The simple structure of the model allows us to highlight the interplay between rigidity and temperature, and to provide insight in the underlying microscopic mechanisms of stress homogenization and diffuse failure. 

\section{\label{sec:mm}Model and methods}

We consider diluted spring networks with a 2D triangular topology consisting of $L \times L$ nodes separated by a distance $\ell_0$. Nearest neighbors are connected by bonds, which gives a maximum network connectivity $z_{\text{max}} = 6$. The network is subsequently randomly diluted by removing a fraction $1-p$ of the bonds, such that the average connectivity becomes $\langle z\rangle = p\,z_{\text{max}}$. Periodic boundary conditions are employed in all directions. The bonds are harmonic (linear) springs with spring constant $\mu$ and rest length $\ell_0$. Excluded volume interactions are not present in the system. During fracture simulations, bonds break irreversibly when their deformation $\Delta \ell$ exceeds a rupture threshold $\lambda$, that is the same for all the springs. We will focus on networks with $\lambda=0.03$.

 Simulations are performed using LAMMPS~\cite{Plimpton1995} and nodes follow Langevin dynamics:
 \begin{equation}
m \frac{d^2 \mathbf{r}}{dt^2}= \mathbf{F} - \zeta \frac{d\mathbf{r}}{dt} + \sqrt{2m \zeta k_B T  }R(t) \;\;, 
 \end{equation}
 where $\mathbf{F}=\mu \Delta \mathbf{\ell}$, $m$ is the mass that is set to unity, $\zeta$ is the friction coefficient related to the (implicit) solvent viscosity, $k_B$ the Boltzmann's constant, $T$ the temperature and $R(t)$ white noise with zero-mean. The integration time step is set to $\delta t=0.001 \tau$, where $\tau=\sqrt{m\ell_0^2/\mathcal{E}}$ is the unit time of our simulations, and the energy scale is set to $\mathcal{E}=1$. The spring stiffness (in reduced units) is set to $\mu=1000$. In our analysis, we will use the reduced temperature $T^*=k_{\mathrm{B}}T / \mu\ell_0^2$, indicating the ratio between thermal and elastic energies. We fix $\zeta=10$ (by setting the damping factor in LAMMPS to $0.1$). In addition, we can define a relaxation time scale $\tau_\text{relax}=\zeta/\mu$, the time that a node requires to travel a distance $\ell_0$ if it is subjected to a force $\mu\ell_0$. For our typical set of simulations $\tau_\text{relax}=5\cdot10^{-3}~\tau$.
 
\subsection{Measuring linear modulus and non-affinity}
We calculate the linear Young's modulus $E$ from the difference in average stress at 0\% strain and 1.5\% strain. At both strain values the system is equilibrated for 100$\tau$ and averaged over 1900$\tau$. The stress $\sigma$ is defined as the $yy$-component (along the deformation axis) of the virial stress tensor normalized by $\mu$. We also calculate the non-affinity parameter (at 1.5 \% strain) defined as

\begin{equation} \label{eq:NAMech}
    \Gamma_{\text{mech}}= \frac{\langle \left( \overline{\mathbf{r}} - \overline{\mathbf{r}}_\text{aff} \right)^2\rangle}{\varepsilon^2}\;\;, 
\end{equation}
where $\overline{\mathbf{r}}$ is the time averaged position of the individual nodes after an applied deformation of 1.5\% strain and $ \overline{\mathbf{r}}_\text{aff}$ the position assuming an affine displacement of 1.5\% strain with respect to the time averaged position of individual nodes at rest. $\varepsilon$ is the strain and $\langle . \rangle$ indicates the average over all nodes. For these simulations, both non-percolating clusters and primary dangling ends (i.e. nodes that are only connected to one bond after dilution) are iteratively removed from the network.
  
\subsection{Non-linear elasticity and fracture}
The network is uniaxially deformed in the $y$-direction up to 100\% strain with a fixed strain rate of $\dot{\varepsilon}=0.001 \tau^{-1}$ (i.e. $0.0001\%$ deformation per unit time) while the lateral size is kept constant. We remap the node positions between time steps, temporarily enforcing affine deformation. The deformation is relatively slow compared to the relaxation time,  $\dot{\varepsilon}=5\cdot 10^{-6} \tau_\text{relax}^{-1}$, which indicates the system has time to respond to the affine deformation via structural rearrangements. Results for varying $\dot{\varepsilon}$ are reported in the ESI. A quick equilibration run of 50$\tau$ precedes the deformation. The network response is quantified by looking at the stress $\sigma$ as a function of strain $\varepsilon$. In addition, we follow the instantaneous non-affine response
 
\begin{equation} \label{eq:NAInst}
    \Gamma_= \frac{\langle \left( \mathbf{r} - \overline{\mathbf{r}}_\text{aff} \right)^2\rangle}{\varepsilon^2}\;\;, 
\end{equation}
 
\noindent where the affine response is calculated with respect to the equilibrium position of the nodes at 0.0\% strain (averaged over 1900$\tau$). Please note that $\Gamma$ is only based on the instantaneous positions and the non-affine response is therefore a combination of both rigidity controlled non-affine network rearrangements and instantaneous thermal fluctuations (see ESI for details on the relation between $\Gamma$ and $\Gamma_{\text{mech}}$).

In fracture simulations bonds are broken every 100 steps (i.e. 0.1$\tau$) when the connected nodes are separated by a distance more than $\ell_0+\lambda$. From the measured stress-strain curves we extract several quantities (see Fig.~\ref{fig:introfig}). All quantities are averaged over several configurations and expressed in reduced units. The peak stress $\sigma_p$ is defined as the highest measured stress, and the peak strain $\varepsilon_p$ is its corresponding strain value.  The maximum stress drop $\Delta\sigma_{\text{max}}$ is calculated according to a procedure~\cite{Bonfanti2018,Dussi2020} where we i) calculate the derivative of the stress-strain curve, ii) make a list of consecutive data points which have a negative derivative and note the initial and final strain of each interval, iii) calculate the stress drops by subtracting the stress at the final strain from the stress at the initial strain, iv) identify the largest stress interval, which corresponds to maximum stress drop  $\Delta\sigma_{\text{max}}$. For the stress distribution analysis, we make instantaneous histograms of the bond lengths $\ell_i$ during the simulation at every percent strain. Based on these histograms, we calculate the excess kurtosis
\begin{equation}
\kappa_e=\frac{\sum_{i}(\ell_i - \langle\ell\rangle)^4}{N_b \, s^4} -3 \;\; ,
\end{equation}
where $s$ is the standard deviation of the histogram, $N_b$ the total number of bonds, and $\langle \ell\rangle$ the average bond length.  For all these parameters the standard error is calculated as the standard deviation divided by the number of sampled configurations (standard error of the mean). The errors are shown when they are larger than the symbols displayed in the graphs.
 
\section{\label{sec:res}Results and discussion}
Using Langevin Dynamics simulations, we uniaxially deform diluted triangular central-force spring networks to study both linear and non-linear network mechanics. By analyzing the network response on both a macroscopic and microscopic level, we gain insight into the effects of thermal fluctuations on the fracture process of networks as well as how the impact of thermal fluctuations is controlled by the network rigidity.

\subsection{\label{sec:lin}The effect of thermal fluctuations on linear elasticity}

First, we study the effects of thermal fluctuations on the linear elasticity.

\begin{figure}[!ht]
\centering
\includegraphics{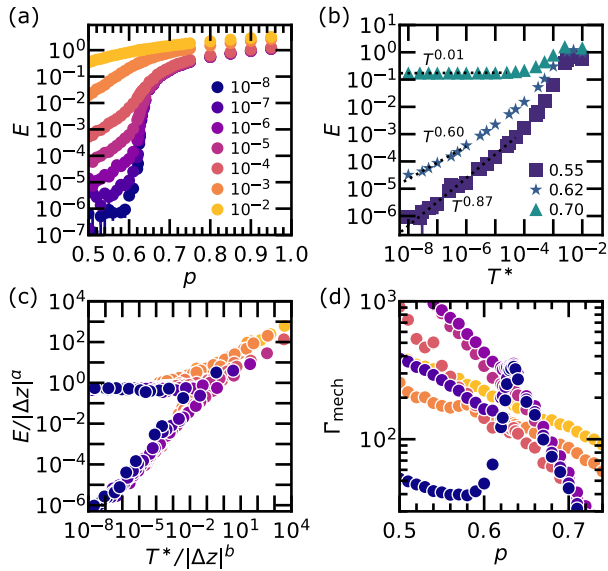}
\caption{\label{fig:linear} Characterization of the linear elastic response for diluted triangular networks of fixed system size $L=128$. (a) The Young’s modulus $E$ as a function of the connectivity parameter $p$ for different temperatures $T^{*}$. (b) Temperature dependence of $E$ for networks below, around, and above the isostatic point (value of $p$ indicated in the legend). The dashed lines indicate the power-law fit $T^{\alpha}$. (c) Rescaling of the Young’s modulus according to Ref.~\cite{Dennison2013a} with $a=1.4$, $b=2.8$ and $z_c=3.78$. (d) The non-affinity parameter $\Gamma_{\text{mech}}$ at 1.5\% strain as a function of $p$ for different temperatures, same legend as (a). Every data point is based on simulations of at least 10 independent configurations.}
\end{figure}

In Fig.~\ref{fig:linear}(a) we plot the linear modulus $E$ of the network as a function of the network connectivity factor $p$ for several reduced temperatures $T^*$. The linear modulus $E$ describes the resistance of a network to deformation and we observe that networks below the isostatic point of mechanical stability~\cite{Maxwell1864} (i.e. networks below $p_{\text{iso}} \approx 0.66$) display a finite linear modulus $E$, which would be absent for athermal systems (in the limit of $T=0$). This finite $E$ is an effect of entropic stiffness, a temperature-dependent phenomenon. As reported in literature~\cite{Dennison2013a,Zhang2016}, the scaling of the linear modulus with temperature $E\propto T^\alpha$ depends on both connectivity and temperature itself. By plotting $E$ as a function of $T^*$ we extract the scaling exponent $\alpha$ from a power-law fit for three different values of $p$, as shown in Fig.~\ref{fig:linear}(b). For a sub-isostatic network with e.g. a connectivity parameter $p=0.55$ the linear modulus scales with $\alpha=0.84$, which roughly corresponds to the dependence found in the anomalous regime as defined in Ref.~\cite{Dennison2013a}, where a shear deformation was instead considered. It was argued that the disordered network structure causes this sub-linear dependence. Whilst there is a clear dependence of the linear modulus on the temperature below the isostatic point, the curves for the different temperatures start to converge when approaching a structurally rigid network (Fig.~\ref{fig:linear}(a)). Accordingly, stiff networks display temperature insensitivity ($\alpha \approx 0$), as can be seen for a network with $p=0.70$ in Fig.~\ref{fig:linear}(b).  As $T^*$ increases, however, the network connectivity becomes less important as the energetic contribution arising from the structural rigidity becomes negligible compared to the entropic elasticity. This is noticeable in Fig.~\ref{fig:linear}(a) where the curve for $T^*=10^{-2}$ is roughly flat for the entire $p$-range, and also in Fig.~\ref{fig:linear}(b) where for $p=0.70$, $E$ increases for $T^*>10^{-3}$. As predicted in Ref.~\cite{Dennison2013a}, we also find a different scaling for networks close to the isostatic point, see e.g. curve for $p=0.62$ in Fig.~\ref{fig:linear}(b). Although the exponent $\alpha$ is slightly different from the findings of Ref.~\cite{Dennison2013a} (where shear deformation and different simulation methods were employed), we were also able to obtain critical rescaling as shown in Fig.~\ref{fig:linear}(c). We can conclude that there are different regimes of dependence for the linear modulus on the temperature based on both rigidity and temperature.

Furthermore, we find similar rigidity dependent behaviour of the thermal fluctuations in the non-affinity parameter $\Gamma_{\text{mech}}$ (Eq.~\ref{eq:NAMech}), reported in Fig.~\ref{fig:linear}(d) as a function of $p$ for different $T^*$. The non-affinity of the network describes how much the time-averaged local deformation differs from the global (externally imposed) deformation. At low $T^*$ we find a peak in non-affine deformation around the isostatic point ($p\approx0.66$). This peak arises from the tendency of the spring network to minimize internal stress upon deformation. If the spring network is far below the isostatic point, the stress can be reduced significantly by a small amount of non-affine rearrangements while at the isostatic point many non-affine rearrangements are required. At the isostatic point, an increase in $T^*$ decreases $\Gamma_{\text{mech}}$, which suggests that thermal fluctuations act as a stabilizing field, similar to the bending rigidity in fiber networks~\cite{Broedersz2011}. However, we note that the effect of thermal fluctuations is always present, even without external deformations, leading to structural rearrangements in the rest state (see ESI). 
Above the isostatic point, we observe that the non-affinity converges for most values of $T^*$ (see ESI for details), which indicates that above the isostatic point the network rigidity dominates the non-affine response. Only if $T^*>10^{-4}$, we see that thermal fluctuations affect the non-affine response, increasing $\Gamma_{\text{mech}}$. This is in contrast to fiber networks, where the non-affinity decreases with an increase in bending rigidity. We hypothesize that this difference occurs because in the case of fiber networks the fibers have a preference to remain straight to minimize stress caused by fibre bending, while in the case of thermal fluctuations an affine displacement of the nodes will not minimize the stress caused by the randomly oriented thermal fluctuations.
Below the isostatic point, the effect of thermal fluctuations on the non-affine response is significant. We observe that at $T^*=10^{-8}$ the non-affine response is the smallest and that a moderate increase in $T^*$ up to $T^*=10^{-6}$ leads to an increase in the non-affine response, corresponding to what is observed for fiber networks. However, we also observe a decrease in $\Gamma_{\text{mech}}$ if the temperature is increased beyond $T^*=10^{-6}$, which is not observed in fiber networks. It is unclear if this deviation is caused by a fundamental difference between thermal fluctuations and bending rigidity as a stabilizing field or that longer equilibration times are required to gain quantitative information on the non-affine response in this regime (see ESI for details). 

In general we find that at a global level thermal fluctuations act as a stabilizing field, dampening rigidity dependent behaviour around the isostatic point. However, our results suggest that the random nature of the thermal fluctuations causes significant differences in the local response with respect to stabilizing fields in athermal systems such as bending.

\subsection{\label{sec:nlin}The effect of thermal fluctuations on non-linear elasticity}

\begin{figure}[!ht]
\centering
\includegraphics{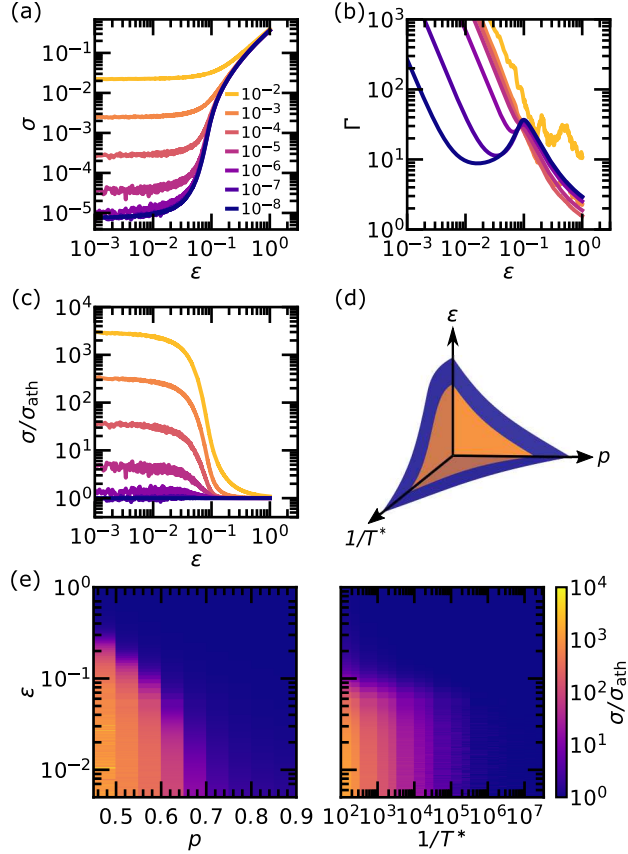}
\caption{\label{fig:nonlinear} The role of temperature on the non-linear elasticity of unbreakable spring networks with fixed system size $L=128$. (a) Stress-strain curve for $p=0.56$ and several $T^*$ (indicated in the legend). (b) Instantaneous non-affinity as a function of strain. (c) The stress ratio $\sigma/\sigma_{\text{ath}}$ as a function of strain for different reduced temperatures $T^{*}$, with $\sigma_{\text{ath}}$ the stress measured in the athermal limit (practically, for $T^{*}=10^{-8}$). (d) Schematic mechanical phase diagram based on the stress ratio with two regimes: a temperature dominated regime (orange) and a mechanically dominated regime (blue). The gradual transition between the regimes depends on deformation $\varepsilon$, network connectivity $p$ and temperature $T^{*}$. (e-f) Two cross-sections of the diagram based on the simulations: (e) $\varepsilon-p$ plane for $T^{*}=10^{-3}$ and (f) $\varepsilon-1/T^*$ plane for $p=0.56$. Every data point is based on simulations of at least 10 independent configurations.}
\end{figure}

In the previous section, we have shown that thermal fluctuations rigidify sub-isostatic networks. Here, we analyze the non-linear elasticity of unbreakable networks to quantify the effect of temperature when networks become more and more strained. The strain-stiffening observed for sub-isostatic networks in the athermal limit has been extensively studied~\cite{Onck2005,Sharma2015,Zagar2015,Feng2016,VanOosten2016,Sharma2016,Bouzid2018,Shivers2019a}. In Fig.~\ref{fig:nonlinear}(a), we report the stress-strain curves for a network with $p=0.56$ at different temperatures. It is evident that the network strain-stiffens for all the $T^*$ investigated. We observe that the onset of strain-stiffening is barely dependent on temperature. Furthermore, the stress response becomes independent of $T^*$ at high strains, similarly to what has been observed for other stabilizing fields, e.g. bending~\cite{Sharma2015,Feng2016}. 

Signatures of strain-stiffening can also be observed in the non-affine response of the network. In Fig.~\ref{fig:nonlinear}(b), we report the instantaneous non-affinity parameter $\Gamma$, that intrinsically includes both the non-affine contributions from instantaneous thermal fluctuations and structural rearrangements. As a result, high non-affinity values can be observed at low strains, where the size of the non-affine thermal fluctuations is large compared to the applied strain. At low temperatures, a peak can be observed in the non-affine response around the onset strain. At high temperatures, this peak is overshadowed by the non-affine thermal fluctuations. At high strain, the network elasticity is controlled by stretching of the bonds and the network response becomes increasingly affine for most temperatures. Only at $T^*=10^{-2}$, the non-affine fluctuations are still visible.

To disentangle the effects of temperature and network connectivity, we normalize the stress-strain curve with the ones obtained in the athermal energy-dominated limit. In particular, we plot the stress ratio $\sigma/\sigma_{\text{ath}}$ in Fig.~\ref{fig:nonlinear}(c), where we used the data obtained at $T^*=10^{-8}$ for $\sigma_{\text{ath}}$. A ratio of $\sigma/\sigma_{\text{ath}} \approx 1$ implies that the mechanical behaviour is basically insensitive to variations in temperature. As can be seen in Fig.~\ref{fig:nonlinear}(c) for a network with $p=0.56$, there is a regime of strain in which the stress ratio depends on $T^*$ that decreases upon stretching the network more and more. At increasing temperature, this stress ratio is both higher at the start and approaches temperature insensitivity at a higher strain. The start of decrease in stress ratio for all temperatures occurs at approximately the same strain value, corresponding to the onset of strain-stiffening. This transition could therefore be interpreted as a transition between a regime dominated by thermal fluctuations to a regime dominated by bond stretching. This is analogous to the bending-to-stretching transition observed in fiber networks~\cite{Onck2005,Buxton2007,Sharma2015}. We summarize these observations in a mechanical phase diagram sketched in Fig.~\ref{fig:nonlinear}(d), where we can distinguish two regimes: a mechanically-dominated regime (blue) where structural rigidity overpowers the effect of thermal fluctuations and a temperature-controlled regime (orange) where thermal fluctuations play a more important role in the elastic behaviour. The transition between these regimes depends on the reduced temperature $T^*$ (and therefore both on the actual temperature $T$ and the bond stiffness $\mu$), the connectivity parameter $p$ and the strain $\varepsilon$. This transition is in general very gradual as can be seen from the two cross-sections of the mechanical phase diagram reported in Fig.~\ref{fig:nonlinear}(e-f), where we show the stress ratio obtained by some of our simulations. When $T^*$ is fixed (Fig.~\ref{fig:nonlinear}(e)) and we increase $p$, we observe a steep decrease in the strain associated to the thermal-stretching transition. Above the isostatic point, the mechanics of the rigid networks is barely affected by thermal fluctuations at this temperature. In Fig.~\ref{fig:nonlinear}(f), we observe that with increasing temperature the stress ratio increases but the strain characterizing the transition seems to reach a limiting value. This limiting value is a result of the onset of strain-stiffening, which is independent of temperature and corresponds to the transition to the elastic regime.

In summary, we identified a rigidity-dependent transition between two regimes where thermal fluctuations are or are not important. In the following sections, we will investigate whether this underlying transition also influences the fracture of these elastic networks. 

\subsection{\label{sec:macro}The effect of thermal fluctuations on macroscopic fracture}

\begin{figure}[!ht]
\centering
\includegraphics{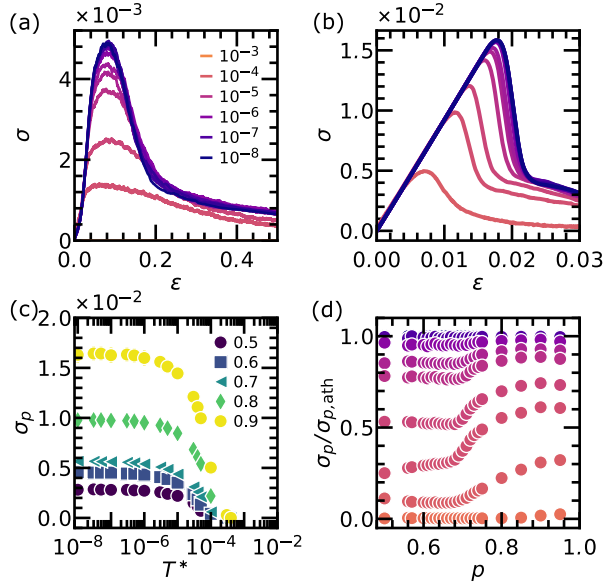}
\caption{\label{fig:macro} Stress-strain curves from fracture simulations for networks with $\lambda=0.03$, $L=128$ at different $T^*$ (see legend) and (a) $p=0.65$, (b) $p=0.90$. Average over 10 independent configurations. (c) Temperature dependence of the peak stress $\sigma_p$ for networks with different connectivity $p$ (values indicated in the legend). Every data point is based on simulations of at least 60 independent configurations. (d) Connectivity dependence of $\sigma_p$ normalized by the peak stress in the athermal limit $\sigma_{p,\text{ath}}$ for several temperatures, same color code as panel (a). Every data point is based on simulations of at least 10 independent configurations. }
\end{figure}

Under athermal conditions, bonds break only after the onset strain, as only at this stage the bonds are under tension. Furthermore, Fig.~\ref{fig:nonlinear} indicates that the contribution of the thermal fluctuations to the stress in the system is significantly reduced beyond the onset strain. Does this mean that there is only a minor influence of temperature on the failure response?

We first focus on macroscopic descriptors and characterize the stress-strain curves obtained from fracture simulations. In Fig.~\ref{fig:macro}(a-b), we show the response of two representative networks with small rupture threshold $\lambda=0.03$ and different connectivity at several temperatures $T^*$. For the network with $p=0.65 \simeq p_{\text{iso}}$ (panel a), a clear decrease in peak stress $\sigma_p$ for increasing $T^*$ is observed, while a variation in the peak strain $\varepsilon_p$ is less evident as the fracture becomes more ductile and the decrease in stress after the peak is less pronounced. For the very rigid network ($p=0.90$, panel b), the decrease in both $\sigma_p$ and $\varepsilon_p$ is clearly observed. Similarly, the fracture becomes more ductile for higher $T^*$, even though a clear stress drop is still recognizable at the highest temperature simulated. In both cases, the networks become weaker with increasing $T^*$. Furthermore, when approaching the athermal limit ($T^* \rightarrow 0$) the peak stress becomes less sensitive to variation in temperature. In Fig.~\ref{fig:macro}(c), we show the temperature dependence of $\sigma_p$ for several connectivities with $\lambda=0.03$. The common trend is little variation at low temperatures, almost a plateau that is indicative of approaching the athermal limit, followed by a decrease when temperature is increased, with $\sigma_p$ eventually dropping to zero when a temperature of $T^* \simeq 10^{-4}$ is reached. On one hand, for low $T^*$ the peak stress is evidently controlled by the network rigidity, as previously investigated in the athermal limit~\cite{Zhang2016, Dussi2020}. On the other hand, when the thermal energy is of the order of $\frac{1}{2}\mu (\lambda \ell_0)^2$ the network structure is irrelevant, as springs spontaneously break and the system shows melting behaviour. We will later describe the melting point using the reduced quantity $k_{\textrm{B}}T/[\frac{1}{2}\mu(\lambda\ell_0)^2] = T^*/(\frac{1}{2}\lambda^2)$. In between these limits, there is a cross-over regime. To better assess the role of rigidity in this intermediate regime, we normalize $\sigma_p$ by its value in the athermal limit $\sigma_{p,\text{ath}}$ and plot this ratio in Fig.~\ref{fig:macro}(d). The transition between the athermal limit, where $\sigma_p/\sigma_{p,\text{ath}}=1$, and the melting limit, where such a ratio goes to zero, depends on a subtle coupling between connectivity and temperature itself. Far below ($p<0.60$) and far above the isostatic point ($p>0.80$) the connectivity plays a small role since at every temperature the curve exhibits two plateaus (at small and large $p$). However, around the isostatic point rigidity and thermal fluctuations are coupled, since at all the intermediate temperatures we can observe a sharp increase in $\sigma_p/\sigma_{p,\text{ath}}$ upon increasing $p$, connecting the two limiting plateaus. On passing, we note that the plateau for small $p$ is lower, suggesting that temperature starts to affect failure of very diluted networks earlier than for networks with large $p$. Furthermore, we speculate that the complex temperature-dependence around the isostatic point arises from locally floppy regions that are rigidified by thermal fluctuations (whose magnitude depends on temperature itself) and are therefore able to sustain and concentrate stress, and break. Since the isostatic point marks the onset of mechanical stability, such effect is largest for networks close to it.

\subsection{\label{sec:micro}The effect of thermal fluctuations on microscopic fracture}

\begin{figure}[!ht]
\centering
\includegraphics{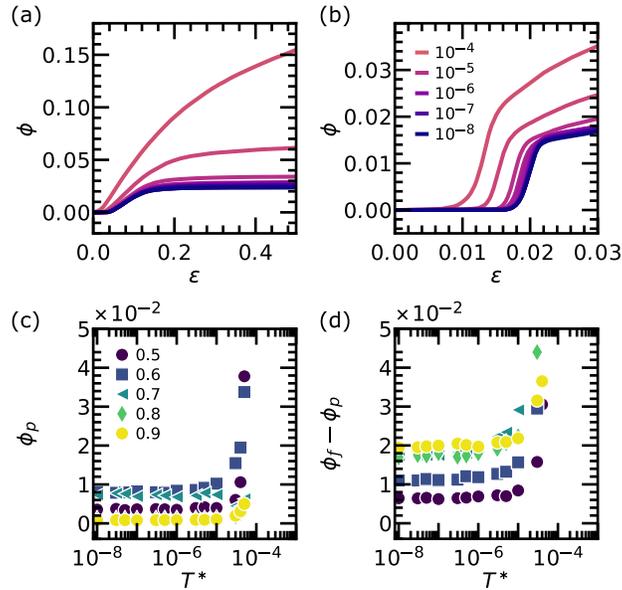}
\caption{\label{fig:micro2} Effect of temperature on the development of microscopic damage. (a-b) Fraction of broken bonds $\phi$ as a function of strain for networks with $L=128$, $\lambda=0.03$ and (a) $p=0.65$, (b) $p=0.90$. Average over 10 independent configurations. (c) Fraction of broken bonds at the peak strain $\phi_p$ (including the peak event) as a function of $T^*$ for a range of dilution factors $p=0.50-0.90$. Every data point is based on simulations of at least 60 independent configurations. (d) Fraction of broken bonds after the peak strain up to failure of the entire system as a function of $T^*$ for a range of dilution factors $p=0.50-0.90$. Data are only shown for systems that lose percolation during the simulations (before 100\% strain). }
\end{figure}

Clearly, the failure response is temperature dependent across the entire rigidity range, but the influence of temperature indeed seems to depend on the distance with respect to the isostatic point. Are these differences also apparent at the microscopic level?
To investigate this, we monitor the number of broken bonds during the simulations. As shown in Fig.~\ref{fig:micro2}(a-b), the fraction of broken bonds $\phi$ as a function of deformation indicates that higher temperature leads to earlier and overall increased damage. However, the effect of temperature is more significant close to the melting temperature $T^* \simeq \frac{1}{2}\lambda^2= 4.5 \cdot10^{-4}$, whereas for lower temperatures the system response is still very much influenced by rigidity. By focusing on the fraction of bonds broken at the peak strain $\phi_p$, counting also the bonds broken during the peak event, as shown in Fig.~\ref{fig:micro2}(c), the diverging behaviour when approaching melting is evident. This increase in broken bonds could explain the decrease in material strength $\sigma_p$. Also the fraction of bonds that break above $\varepsilon_p$ (Fig.~\ref{fig:micro2}(d)), the post-peak response, increases close to the melting point, which points towards a prolonged post-peak response, i.e. higher ductility.

\begin{figure}[!ht]
\centering
\includegraphics{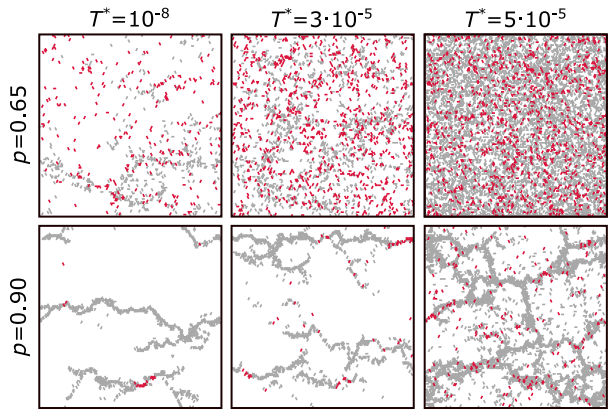}
\caption{\label{fig:micro3} Failure patterns are presented as snapshots of the networks ($L=128$) in their rest state, only showing broken bonds. The bond color indicates whether the bond was broken before (red) or after (grey) $\varepsilon_p$. }
\end{figure}

A direct inspection of the simulation snapshots (Fig.~\ref{fig:micro3}) suggests that bonds that break up to the peak strain $\varepsilon_p$ (red bonds) are dispersed more homogeneously throughout the sample at a higher temperature. The snapshots also reveal a big difference in the response to temperature between networks around ($p=0.65$) and far above the isostatic point ($p=0.90$). Around the isostatic point, the damage up to $\varepsilon_p$ is already diffusive in the athermal limit, and its delocalization is enhanced when the temperature is increased. In contrast, the failure response far above the isostatic point shows a clear transition from crack nucleation in the athermal regime to a more diffuse failure response close to the melting point. However, the post peak response at $p=0.90$ is clearly still dominated by the propagation of cracks. Nevertheless, at high temperatures we observe the development of multiple cracks, sometimes even not perpendicular to the deformation direction, and evidence of crack merging.

In summary, we show that an increase in temperature leads to an increase in diffuse failure, implying suppression of stress concentration before the peak stress. These observations suggest that thermal fluctuations are responsible for two apparently contrasting effects: on the one hand, they create "instantaneous defects" resulting in more regions with broken bonds, that reduce material strength; on the other hand, the fluctuations allow to delocalize stress away from such defects, delaying the propagation of large cracks. As a result, the damage pattern is diffuse throughout the system.

\subsection{\label{sec:macro2}Thermal fluctuations increase the length scale of stress redistribution}

\begin{figure*}[!ht]
\centering
\includegraphics{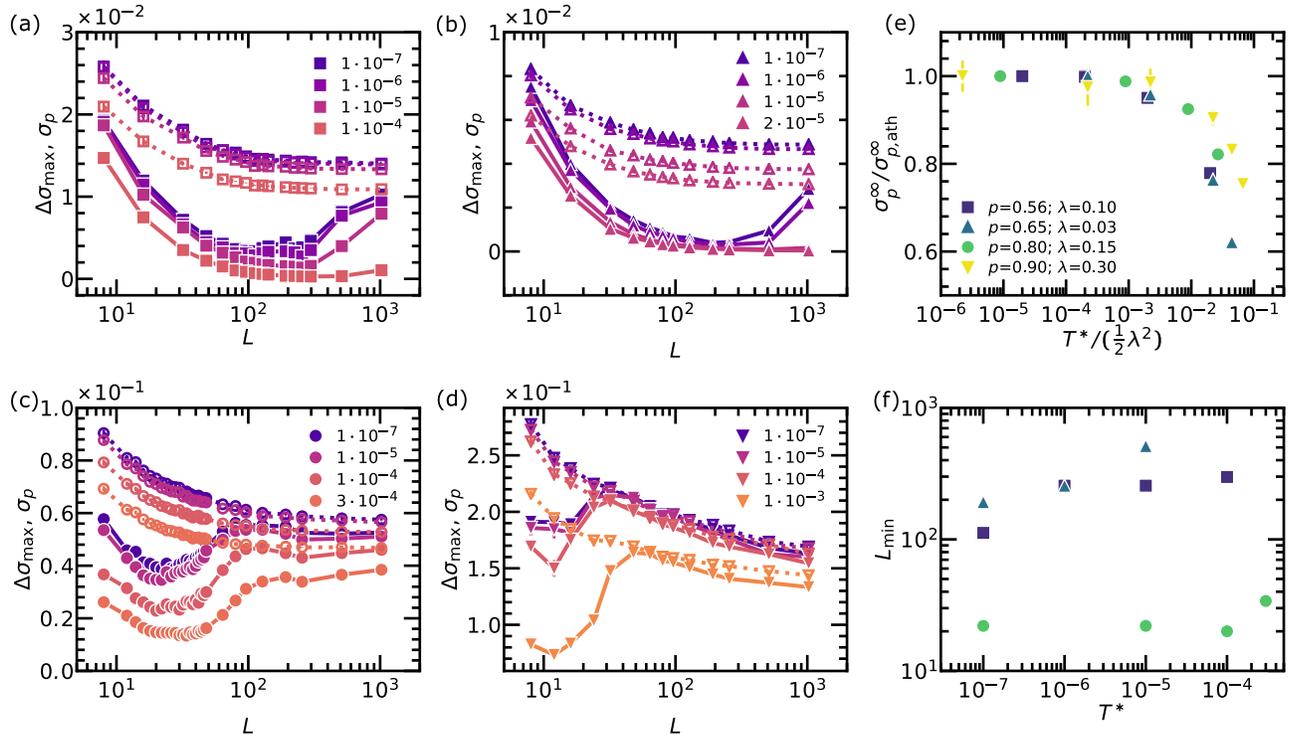}
\caption{\label{fig:SizeDependence} Size-scaling and temperature. (a-d) Maximum stress drop $\Delta\sigma_{\text{max}}$ (closed symbols, solid lines) and peak stress $\sigma_{p}$ (open symbols, dotted lines) as a function of system size $L$ for systems with (a) $p=0.56$ and $\lambda=0.10$ (square), (b) $p=0.65$ and $\lambda=0.03$ (triangle), (c) $p=0.80$ and $\lambda=0.15$ (circle), and (d) $p=0.90$ and $\lambda=0.30$ (downward triangle). Depending on system size the minimum number of independent simulations per data point is 60 ($L=8\ldots128$), 30 ($L=192\ldots256$), 10 ($L=512$) or 5 ($L=1024$). (e) Peak stress in the thermodynamic limit $\sigma^{\infty}_{p}$ normalized by the corresponding value in the athermal limit $\sigma^{\infty}_{p,\text{ath}}$, as a function of reduced temperature $T^{*}$ normalized by its melting value $\frac{1}{2}\lambda^2$. Error bars represent the standard error in the fit for $\sigma^{\infty}_{p}$. (f) Estimate of the system size where $\Delta\sigma_{\text{max}}$ is minimal as a function of $T^{*}$. In both (e) and (f) the marker shape corresponds to the value of $p$ as introduced in panels (a-d).}
\end{figure*}

It is possible that some characteristic lengthscale associated to stress concentration (or equivalently to stress delocalization) exists. For example, we have recently~\cite{Dussi2020} shown that in athermal systems, brittle (abrupt) fracture always occurs for networks above a certain system size $L^*$. This critical size can be tuned by the network rigidity, i.e. by varying $p$ and $\lambda$. The onset of the size-induced brittleness can be determined by looking at the non-monotonic size-dependence of $\Delta \sigma_{\text{max}}$. Here we investigate whether temperature affects this critical system size, which can be interpreted as a characteristic lengthscale over which stress concentrates. This analysis is therefore another way to further assess the role of temperature on stress concentration.

Therefore, we examine how thermal fluctuations affect the macroscopic fracture descriptors for different system sizes, focusing on the maximum stress drop $\Delta \sigma_\text{max}$ that quantifies fracture abruptness. In Fig.~\ref{fig:SizeDependence}(a-d), we plot the size-scaling of the maximum stress drop $\Delta \sigma_{\text{max}}$ (closed symbols) together with the peak stress $\sigma_p$ (open symbols) for four combinations of $p$ and $\lambda$ at different temperatures. In all cases, we observe a monotonic decrease of $\sigma_p$ as a function of the system size. These trends can be fitted by a power law $\sigma_p = (L/\alpha)^{-\beta}+\sigma_{p}^{\infty}$, where $\sigma_{p}^{\infty}$ is the failure stress in the thermodynamic limit (infinite system size), $\beta$ the size scaling exponent and $\alpha$ a fitting constant. In Fig.~\ref{fig:SizeDependence}(e), we plot $\sigma_{p}^{\infty}$ normalized by its athermal value $\sigma_{p,\text{ath}}^{\infty}$ as a function of $T^*$ normalized by $\frac{1}{2}\lambda^2$ (see ESI for the other fitting parameters). The observed trend underlying a transition from low $T^*$ to melting is consistent with the data at fixed system size and fixed rupture threshold $\lambda$ presented earlier in Fig.~\ref{fig:macro}. Here, we can also appreciate the effect of varying $\lambda$ in the intermediate temperature regime. For example, the largest $\lambda=0.30$ (downward triangles, networks with $p=0.90$) shows a steeper decrease in the normalized fracture stress, suggesting that thermal effects kick in at higher temperatures for these very rigid networks.

Finally, we focus on the maximum stress drop $\Delta \sigma_{\text{max}}$. From Fig.~\ref{fig:SizeDependence}(a-d), we observe that a non-monotonic trend is observed in basically all cases, consistent with our previous results in the athermal limit~\cite{Dussi2020}. We speculated that the initial decrease, implying a more ductile fracture upon increasing system size, is associated to the rupture and reformation of locally stressed regions (often consisting of aligned springs, and sometimes called force chains~\cite{Heussinger2007,Arevalo2015,Liang2016,Zhang2016}). However, upon increasing the system size $\Delta \sigma_{\text{max}}$ starts to increase, suggesting that stress concentration is present in the system, since it fractures in a more abrupt way. At even larger $L$, $\Delta \sigma_{\text{max}}$ decreases again, now following the same trend for the peak stress $\sigma_p$ that sets the upper bound to the possible stress drop. In Fig.~\ref{fig:SizeDependence}(c) the entire trend is visible for the system sizes explored in this work, whereas in the other panels only parts of it are captured. Importantly, for all systems, the trend depends on temperature. In particular, in Fig.~\ref{fig:SizeDependence}(f) we quantify the effect of temperature by plotting the system size $L_\text{min}$ corresponding to the minimum $\Delta \sigma_{\text{max}}$ as a function of $T^*$. We observe that thermal fluctuations increase the value of $L_\text{min}$, which can be interpreted as a lengthscale for stress concentration. The role of temperature seems particularly relevant at low connectivity, where the stress is already very delocalized in the athermal limit.

In summary, we find that also in the thermodynamic limit there is a crossover from an athermal regime to a melting regime where the failure behaviour is determined by both rigidity and thermal fluctuations. Moreover, thanks to the analysis of $\Delta\sigma_{\text{max}}$, we find evidence that temperature increases the region over which stress is delocalized. 

\subsection{\label{sec:micro2}Thermal fluctuations homogenize stress}

\begin{figure*}[ht!]
\centering
\includegraphics{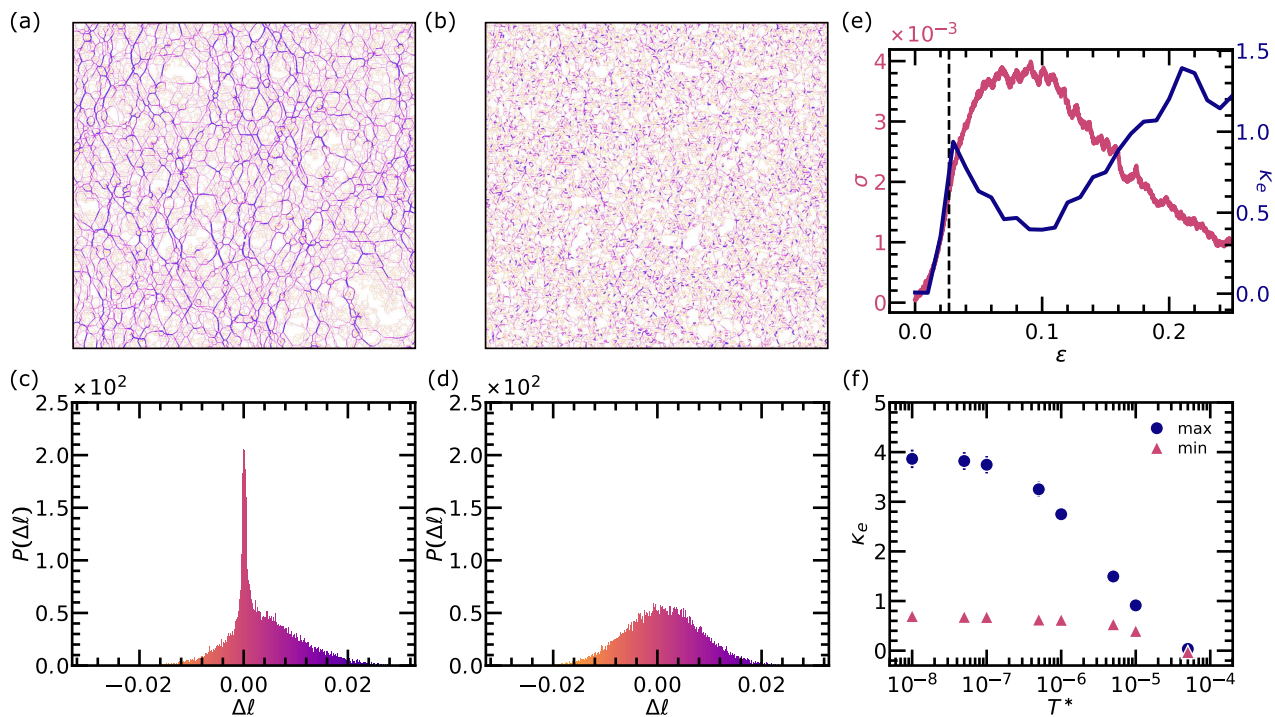}
\caption{\label{fig:micro1} Influence of temperature on the stress distribution in networks with $p=0.65$, $\lambda=0.03$, $L=128$. (a-b) Snapshots of the local bond extension at 10\% strain and (c-d) the corresponding histograms. The color scale indicates the amount of bond extension (i.e. local stress) on the network bonds, purple corresponds to high stress and orange to low stress. (a) For $T^* =10^{-7}$, aligned highly-stressed bonds ("force chains") are visible. (b) For $T^{*}=5\cdot10^{-5}$, the stress distribution is highly homogeneous. (e) Stress and excess kurtosis $\kappa_e$ of the stress histogram, a measure of heterogeneity, as a function of the strain for a single simulation. The dashed black line indicates the strain at which the first bond breaks, also corresponding to the maximum of $\kappa_e$; whereas the minimum of $\kappa_e$ occurs at the peak stress. (f) Maximum and minimum of $\kappa_e$ as a function of $T^{*}$. Every data point is based on simulations of 10 independent configurations.}
\end{figure*}

The delocalization of stress is mediated by structural rearrangements in the network. Therefore, if temperature helps to delocalize stress as suggested by Fig.~\ref{fig:micro3} and Fig.~\ref{fig:SizeDependence}(f), this must be evident in the distribution of stress within the network. 
In Fig.~\ref{fig:micro1}(a-b) we show two snapshots of the same deformed network ($p=0.65$) at two different temperatures, together with the associated histogram of the bond deformation\ref{fig:micro1}(c-d), which is equivalent to the probability distribution of the microscopic stresses since the springs are linear. Both networks are deformed up to 10\% strain, which is close to $\varepsilon_p$ and well above the onset strain discussed in Sec.~\ref{sec:nlin}. At the lower temperature, the stress is distributed very heterogeneously, as indicated by an asymmetric distribution with an exponential tail containing few bonds carrying a high load. The regions of high stress, typically composed of aligned highly-stressed bonds, that we call force chains, can be readily identified in the simulation snapshot. On the contrary, for the higher temperature, the distribution is very symmetric, resembling a Gaussian distribution, and the force chains can not be identified. This indicates that even above the onset strain thermal fluctuations act as a stabilizing field as discussed in Sec.~\ref{sec:lin} and do affect the distribution of stress in the network.
To quantify this heterogeneity, we calculate the excess kurtosis $\kappa_e$ of the stress distribution, an indicator of the tail heaviness of a distribution (being zero for a Gaussian). This measure has been recently used to quantify stress heterogeneities in porous materials~\cite{Laubie2017}. To illustrate how the heterogeneity of the stress distributions is linked to the macroscopic stress evolution, in Fig.~\ref{fig:micro1}(c) we plot both $\kappa_e$ and $\sigma$ as a function of strain for an example simulation run. As observed in most cases, the strain-stiffening of the network is accompanied by a similar increase in kurtosis. The stress distribution becomes more heterogeneous until (approximately) the first bond breaks (the dashed black line in Fig.~\ref{fig:micro1}(c)), after which strain softening occurs and the stress distribution becomes more homogeneous. This decrease in heterogeneity is presumably caused by redistribution of the stress after bonds are broken. Strikingly, in correspondence with the peak stress, a local minimum for the kurtosis is observed. The subsequent stress drops are instead accompanied by an increase in $\kappa_e$, and therefore in the microscopic stress heterogeneity. This increase indicates stress concentration somewhere in the network leading to significant bond breakage that does not allow for larger stress response. To show how the stress heterogeneity changes with temperature, we plot the maximum and the minimum of $\kappa_e$ as a function of $T^*$ in Fig.~\ref{fig:micro1}(d). To determine the minimum kurtosis, we take the smallest value of $\kappa_e$ in a strain interval close to the peak strain $\varepsilon_p$, to avoid lower values that might be found before strain-stiffening. Analogously, for the maximum kurtosis, we only look at the maximum up to and including the peak strain, to avoid post-peak values. Both quantities clearly decrease when the temperature is raised, but follow different curves. In particular, the maximum of $\kappa_e$, that is associated to the network strain-stiffening, is immediately sensitive to temperature changes, in line with our previous observations on the non-linear elasticity (Sec.~\ref{sec:nlin}), while the minimum of $\kappa_e$, associated to the fracture peak, exhibits an initial temperature insensitive interval, similarly to the other fracture descriptors investigated above. Furthermore, as the temperature increases, the difference between the maximum and minimum becomes smaller, and eventually both quantities reach zero (homogeneous stress distribution) at the melting temperature. Note that, while we have shown here results only for a given connectivity, the fact that a higher temperature allows for better redistribution of the stress during fracture was a consistent observation in our simulation study.

\section{\label{sec:concl}Summary and conclusions}
In this work we explored the relation between rigidity and network failure under the influence of thermal fluctuations. Our results demonstrate that thermal fluctuations couple with network rigidity and affect the non-linear mechanics of elastic networks. In general, thermal fluctuations lead to a lower failure strength (Fig.~\ref{fig:macro}), an increased ductility and increased fraction of broken bonds (Fig.~\ref{fig:micro2}). We have shown that at the microscopic level the failure response is altered with respect to the athermal case in two ways: i) bond failure can be activated by instantaneous thermal fluctuations, creating additional weak spots, and ii) stress is delocalized, suppressing the expansion of existing defects (Fig.~\ref{fig:micro3}). We reveal that temperature acts as a stabilizing field that resists large structural non-affine deformation within the network (Fig.~\ref{fig:linear} and Fig.~\ref{fig:nonlinear}). Specifically, the thermal fluctuations increase the lengthscale over which stress is redistributed, which can be quantified via the maximum stress drop (Fig.~\ref{fig:SizeDependence}) and the excess kurtosis (Fig.~\ref{fig:micro1}). Although these trends can be observed for all connectivities, there are distinct damage mechanisms above and below the isostatic point. Above the isostatic point, the failure up to the peak stress shifts from crack nucleation at a single site to a more diffuse failure pattern, while around the isostatic point the failure response is already delocalized in the athermal limit and the fraction of broken bonds is enhanced approaching the melting point (Fig.~\ref{fig:micro3}). These distinct failure processes might explain the difference in how the peak strain depends on temperature with respect to rigidity (Fig.~\ref{fig:macro}).

We note that at a first glance elastic networks subjected to thermal fluctuations behave like athermal networks in a stabilizing field. However, the instantaneous nature of the thermal fluctuations introduces important differences. It is striking that, without any applied deformation, the thermal fluctuations induce structural rearrangements of the average network structure (see ESI). Furthermore, providing enough time, the thermal fluctuations allow the failure of bonds even if they are not intrinsically under tension (activated failure), leading to diffuse damage. A final consequence of introducing thermal fluctuations is that time becomes an important parameter. In our simulations the system was deformed at a constant strain rate, i.e. it was driven at a given speed. If the driving speed is too low, the system will melt due to the process of activated failure. If the driving speed is too high, the system has no time for stress relaxation as it is held back by the viscous surroundings. Therefore, the failure response of an elastic network is generally determined by the coupling between the driving speed, viscosity, rigidity and thermal fluctuations. Our study was focused on a regime in which driving and viscosity effects were small (see ESI for discussion). 

This work provides new insight into the relation between the static network structure, thermal fluctuations and the failure response of network materials. Above all, it shows that the ratio $T/\mu$ can be used to significantly change the failure response of a network. In relating this knowledge to soft matter systems, a clear challenge is to better understand the influence of the physics at the element level, such as plastic rearrangements in colloidal gel strands\cite{Verweij2019} and temperature sensitivity of elastic elements such as semi-flexible fibres\cite{Storm2005}.

\section*{Conflicts of interest}
There are no conflicts to declare.

\section*{Acknowledgements}
This work is part of the SOFTBREAK project funded by the European Research Council (ERC Consolidator Grant No. 682782).


\renewcommand\refname{References}

 
\bibliographystyle{unsrt}

\end{document}


\maketitle

\subsection*{ Dependence of non-affinity on the reference coordinates }
\begin{figure}[!ht]
    \includegraphics{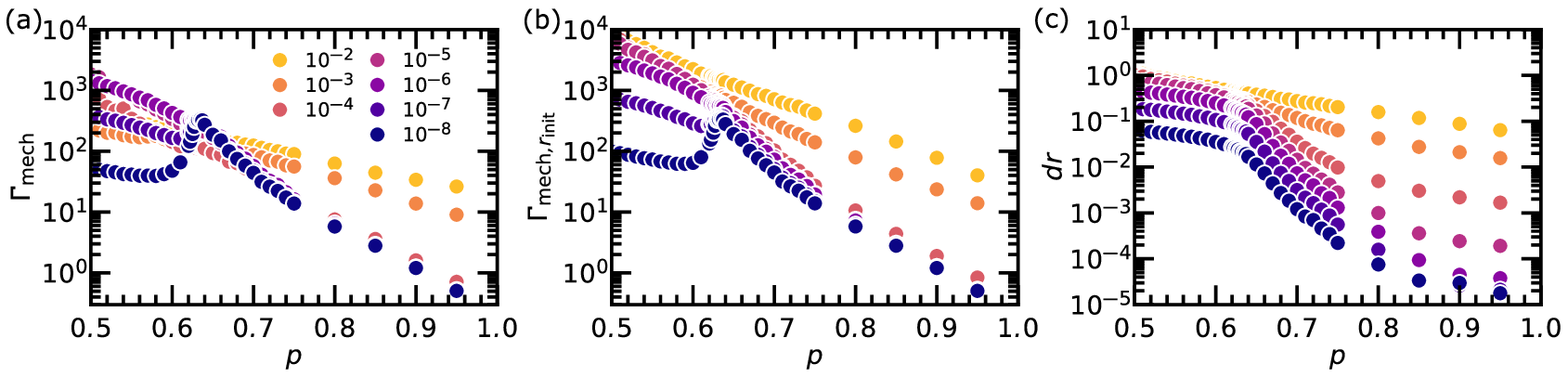}
    \caption{ \label{fig:SINonAffinity} Non-affine response in the linear regime versus $p$ for a range of $T^*$ (colors are indicated in the legend). (a) The time-averaged (over 1900$\tau$) non-affine response at 1.5\% strain with respect to the time-averaged position of the nodes at 0\% strain $\overline{\mathbf{r}}_0$. (b) The time-averaged non-affine response at 1.5\% strain with respect to the coordinates $r_{\text{init}}$ of the initial configuration (a regular triangular lattice). (c) The ensemble averaged displacement $dr = \langle |\overline{\mathbf{r}}_0-\mathbf{r}_{\text{init}}|\rangle$ at 0\% strain. Every data point is based on simulations of at least 10 independent configurations.}
\end{figure}

We calculate the non-affine deformation with respect to the time averaged position at 0\% strain (Fig.~S\ref{fig:SINonAffinity}(a)) $\overline{\mathbf{r}}_0$. Alternatively, the non-affinity can be calculated with respect to the initial positions of the nodes (located on a regular triangular lattice) $\mathbf{r}_{\text{init}}$ (Fig.~S\ref{fig:SINonAffinity}(b)). The difference between Fig.~S\ref{fig:SINonAffinity}(a) and Fig.~S\ref{fig:SINonAffinity}(b) indicates that temperature has an effect on the equilibrium node positions at 0\% strain. This is in contrast with athermal networks and bending stabilized networks, where the equilibrium node positions are equal to the initial position of the nodes. Fig.~S\ref{fig:SINonAffinity}(c) shows that the average displacement of the nodes from their initial position to their equilibrium position $dr = \langle|\overline{\mathbf{r}}_0 -\mathbf{r}_{\text{init}}|\rangle$ depends on both temperature and connectivity. Especially below the isostatic point there are significant reorganizations within the network. However, at high temperatures also the network structure well above the isostatic point is affected. Clearly, the thermal fluctuations do affect the equilibrium structure at 0\% strain.

\subsection*{ Size of the thermal fluctuations}
\begin{figure}[!ht]
    \includegraphics{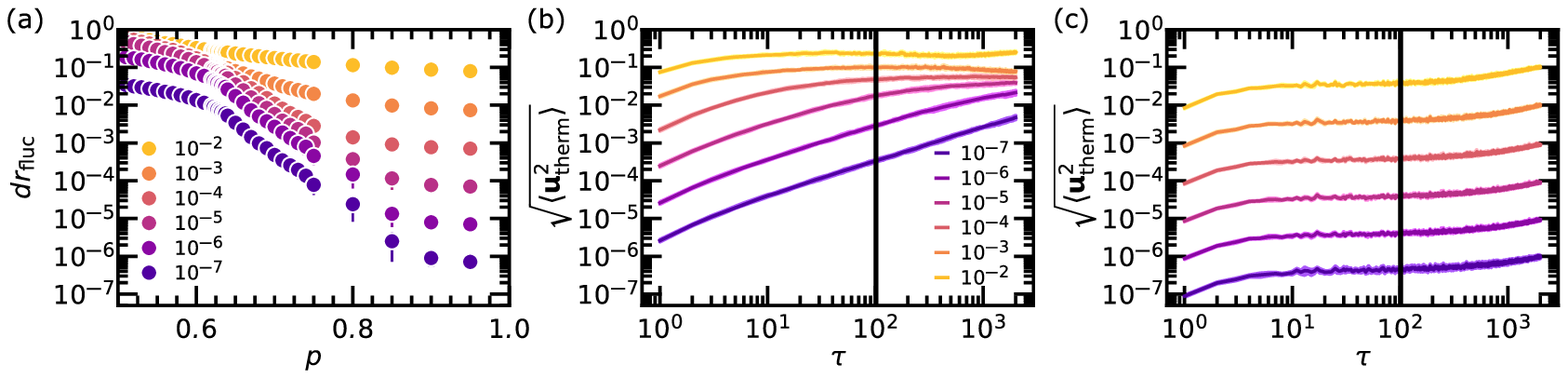}
    \caption{ \label{fig:SIFluctuations} The size of the thermal fluctuations of the node positions. (a) the time-averaged (over 1900~$\tau$) root mean squared fluctuation size $dr_{\text{fluc}} = \sqrt{\overline{\langle \mathbf{u}^2_{\text{therm}}\rangle}}$ versus $p$ for a range of $T^*$ (colors are indicated in the legend). (b-c) The development of $\sqrt{\langle \mathbf{u}^2_{\text{therm}}\rangle}$ as a function of time (2000~$\tau$ in total). The black line is placed at 100~$\tau$, time-averages for $\Gamma_{\text{mech}}$ and $\sqrt{\langle \mathbf{u}^2_{\text{therm}}\rangle}$ are based on data past this line. (b) $p=0.65$ and (c) $p=0.90$. Every data point is based on simulations of at least 10 independent configurations.}
\end{figure}

To quantify the size of the thermal fluctuations, we monitor the root mean squared displacement of the nodes with respect to their equilibrium position $\sqrt{\langle \mathbf{u}^2_{\text{therm}}\rangle}$ and define the size of the fluctuations as $dr_{\text{fluc}}=\sqrt{\overline{\langle \mathbf{u}^2_{\text{therm}}}\rangle}$ with $\overline{\;\cdot\;}$ representing a time-average. From Fig.~S\ref{fig:SIFluctuations}(a) it is clear that the size of the fluctuations of the nodes depends on both temperature and connectivity. In general the size of the fluctuations decreases with an increase in connectivity, indicating there is feedback between the number of constraints imposed on a node in the network and how far the nodes can move. Before measuring $\sqrt{\overline{\langle \mathbf{u}^2_{\text{therm}}}\rangle}$, all systems are subjected to the same calibration run of 100$\tau$ (see Fig.~S\ref{fig:SIFluctuations}(b)/(c)). We note that for $p=0.65$ (Fig.~S\ref{fig:SIFluctuations}(b)) the required time to reach a stable value of $\sqrt{\overline{\langle \mathbf{u}^2_{\text{therm}}}\rangle}$ is longer for lower temperatures, indicating that the rate of equilibration depends on temperature. Furthermore, we see that at the higher connectivity value $p=0.90$ (Fig.~\ref{fig:SIFluctuations}(c)) the fluctuations are smaller and reach their equilibrium value faster.

\subsection*{Relation between time-averaged non-affinity and instantaneous non-affinity}

In an athermal elastic network the position of the crosslinks is determined by the applied deformation and the non-affine response of the nodes. In a thermal elastic network, the positions of the nodes are also influenced by thermal fluctuations of the nodes. The position of a node $\mathbf{r}$ under uniaxial extension $\varepsilon$ can therefore be described as

\begin{equation} \label{eq:pos}
    \mathbf{r} (\varepsilon,T,p) = \overline{\mathbf{r}}_0 + \mathbf{u} _{\text{aff}}(\varepsilon) +\mathbf{u} _{\text{naff}}(\varepsilon,T,p) + \mathbf{u}_{\text{therm}}(\varepsilon,T,p) \;\;,
\end{equation}

\noindent where $p$ is the network connectivity parameter, and $T$ temperature. $\mathbf{u}$ stands for a displacement vector and $\overline{\mathbf{r}}_0$ is the time averaged position at 0\% strain. If a system is fixed at a certain strain $\varepsilon$ the average position of the particle over time will be,

\begin{equation} \label{eq:pos_ave}
    \overline{\mathbf{r}} = \overline{\mathbf{r}}_0 + \mathbf{u}_{\text{aff}}(\varepsilon) + \mathbf{u} _{\text{naff}}(\varepsilon,T,p) \;\;,
\end{equation}

\noindent assuming $|\overline{\mathbf{u}}_{\text{therm}}|=0$. If we instead monitor the average size of the fluctuations of the nodes we find that

\begin{equation} \label{eq:pos_fluc}
    \overline{ \left(\mathbf{r} - \overline{\mathbf{r}} \right)^2} = \mathbf{u}^2 _{\text{therm}}(\varepsilon,T,p)
\end{equation}

\noindent Note that in case of drift in the system center of mass, this needs to be taken into account. In our case, we assume $<\mathbf{r}>=\mathbf{r}_{\text{com}}$. Below, we will detail how these contributions are related to the measure for non-affinity $\Gamma$.

\subsubsection*{Non-affine response of time averaged positions}
We will start with the non-affinity based on time-averaged positions (Eq.~\ref{eq:pos_ave}), as this parameter is directly related to the non-affinity parameter discussed for athermal systems that describes the size of non-affine rearrangements of the network.

\begin{equation}
    \Gamma_{\text{mech}} = \frac{\langle(\overline{\mathbf{r}} – \overline{\mathbf{r}}_{\text{aff}})^2\rangle}{\varepsilon^2} = \frac{\langle(\mathbf{u}_{\text{naff}})^2\rangle}{\varepsilon^2} \;\;.
\end{equation}

\noindent To simplify the equation our definition for $\overline{\mathbf{r}}$ is used (Eq.~\ref{eq:pos_ave}) and the definition $\overline{\mathbf{r}}_{\text{aff}} = \overline{\mathbf{r}}_0 + \mathbf{u}_{\text{aff}}$.

\subsubsection*{Non-affine response of instantaneous positions}
In a system with thermal fluctuations, the instantaneous positions of the node will also be determined by (non-affine) thermal fluctuations (See Eq.~\ref{eq:pos}). While monitoring the non-affinity during a continuous deformation, the non-affinity parameter $\Gamma$ will therefore include both the effects of non-affine rearrangements and thermal fluctuations.

\begin{equation}
    \Gamma = \frac{\langle(\mathbf{r}– \overline{\mathbf{r}}_{\text{aff}})^2\rangle }{\varepsilon^2} = \frac{\langle(\mathbf{u}_{\text{naff}} + \mathbf{u}_{\text{therm}})^2\rangle}{\varepsilon^2}\;\;.
\end{equation}

\subsubsection*{ The relation between $\Gamma$ and $\Gamma_{\text{mech}}$ }
\noindent An essential difference with respect to athermal networks is that $\Gamma$ is a result of both non-affine structural rearrangements and thermal fluctuations. Here we show the relation between $\Gamma_{\text{mech}}$ and $\Gamma$. We can rewrite the non-affinity such that we only have sums over all particles on the right hand side.

\begin{equation}
    [\Gamma - \Gamma_{\text{mech}}] N\varepsilon^2 = \sum((\mathbf{u}_{\text{naff}} + \mathbf{u}_{\text{therm}})^2) – \sum((\mathbf{u}_{\text{naff}})^2) = \sum(   \mathbf{u}_{\text{therm}}^2 + 2\mathbf{u}_{\text{naff}}\cdot\mathbf{u}_{\text{therm}} ) \;\;.
\end{equation}

\noindent Returning to the averages over all particles we can now distinguish a term related to thermal fluctuations and a cross-term related to both the non-affine deformation and the thermal fluctuations.

\begin{equation}
    \Gamma =\Gamma_{\text{mech}} + \frac{\langle \mathbf{u}_{\text{therm}}^2 \rangle}{\varepsilon^2}  + \frac{2\langle \mathbf{u}_{\text{naff}}\cdot\mathbf{u}_{\text{therm}}\rangle}{ \varepsilon^2}
\end{equation}

\noindent Hence, the contributions of structural rearrangements and thermal fluctuations to $\Gamma$ can not be decoupled.

\subsection*{\label{sec:viscrate}The influence of driving and viscosity}

\begin{figure*}[!ht]
\centering
\includegraphics{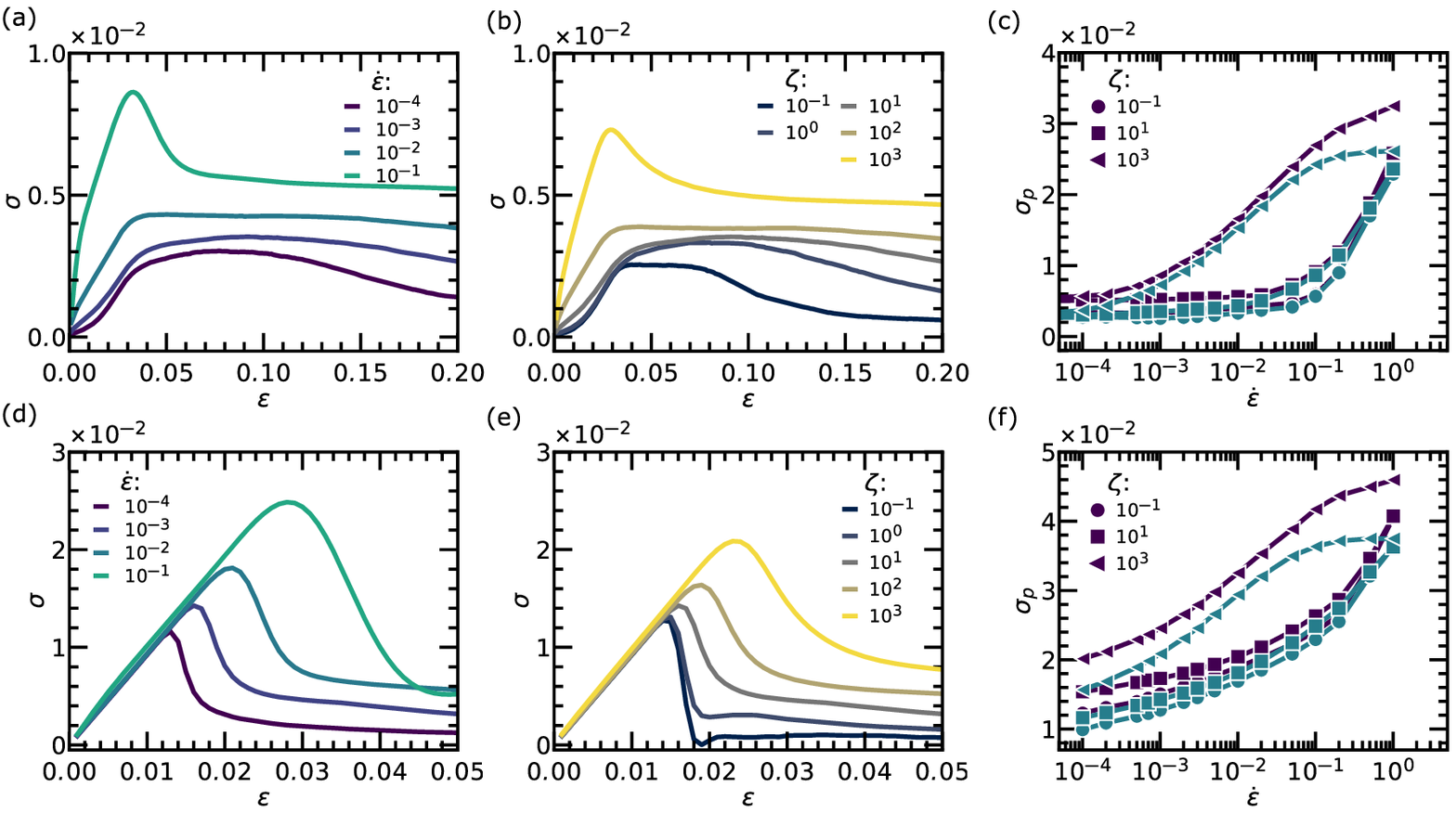}
\caption{\label{fig:ViscosityDriving} The influence of driving and friction coefficient on the stress-strain response. (a) Stress versus strain for a network around the isostatic point ($p=0.65$, $L=1024$, $\zeta=10$, $T^*=1\cdot10^{-5}$) for a range of strain rates (see legend). (b) Stress versus strain for a subisostatic network ($p=0.65$, $L=1024$, $\dot{\varepsilon}=0.001$, $T^*=1\cdot10^{-5}$) for a range of $\zeta$ (see legend). (c) $\sigma_p$ versus $\dot{\varepsilon}$ for a networks with $p=0.65$ and $L=1024$. $\sigma_p$ is determined for $T^*=1\cdot10^{-9}$ (purple) and $T^*=1\cdot10^{-5}$ (blue) the shape of the markers indicates the friction coefficient $\zeta$ (see legend). Stress versus strain for a network far above the isostatic point ($p=0.90$, $L=1024$, $\zeta=10$, $T^*=1\cdot10^{-5}$) for a range of strain rates (see legend). (e) Stress versus strain for a network far above the isostatic point ($p=0.90$, $L=1024$, $\dot{\varepsilon}=0.001$, $T^*=1\cdot10^{-5}$) for a range of $\zeta$ (see legend).  (f) $\sigma_p$ versus $\dot{\varepsilon}$ for networks with $p=0.90$ and $L=1024$. Same color code as in (c). }
\end{figure*}

The mechanical and failure response can depend on both the driving speed and the friction coefficient of the nodes, i.e., viscosity of the implicit solvent. In Fig.~S\ref{fig:ViscosityDriving} we show how these parameters affect the stress-strain curves around the isostatic point (Fig.~S\ref{fig:ViscosityDriving}(a-b)) and far above the isostatic point (Fig.~S\ref{fig:ViscosityDriving}(d-e)). An increase in either $\dot{\varepsilon}$ or $\zeta$ has a similar effect as both parameters affect the relaxation time. In general, an increase in these parameters leads to an increase in ductility. Below the isostatic point, both the pre-peak and post-peak behaviour are affected, while above the isostatic point it is mostly the post peak response. Furthermore, we show the peak stress $\sigma_p$ as a function of the strain rate for $p=0.65$ and $p=0.90$ for a range of $\zeta$ and two values of $T^*$: $1\cdot10^{-9}$ (purple) and $1\cdot10^{-5}$ (blue). We see that typically the peak stress increases with the strain rate. At a higher strain rate, there is less time for stress relaxation, leading to a more affine response and a higher stress in the system. Similarly, we observe that the peak stress increases with an increase in the friction coefficient, which is also related to the time for stress relaxation $\tau_{\text{relax}}=\zeta/\mu$. In the main text, we set the strain-rate $\dot{\varepsilon}=0.001$ and the friction coefficient $\zeta=10$. We can observe that for $p=0.65$ we are in a regime where the effect $\dot{\varepsilon}$ and $\zeta$ on the peak stress is relatively small. At $p=0.90$ the effect of strain rate is bigger, however we do not expect this affects our conclusions.
\newpage

\begin{figure}[!ht]
    \includegraphics{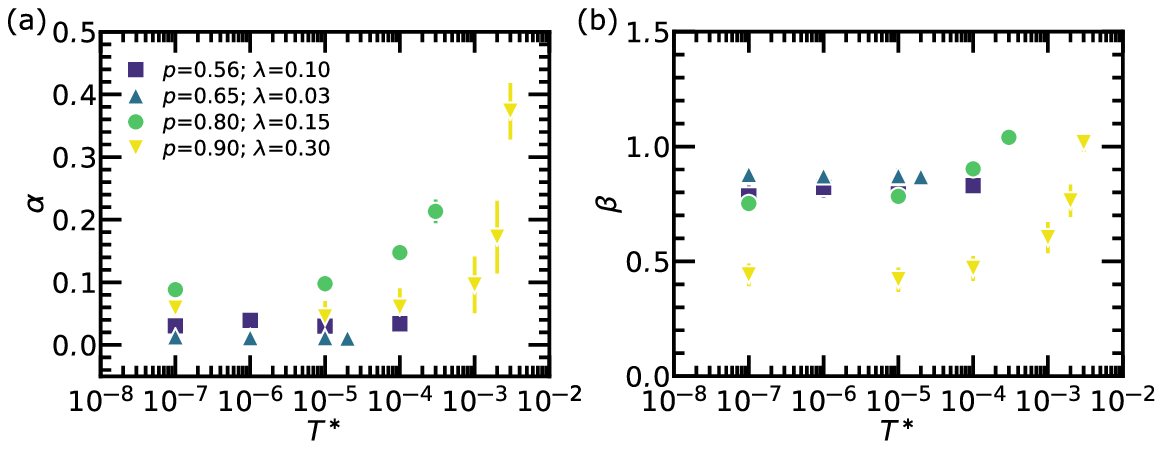}
    \caption{ \label{fig:SIFitting} Fit parameters of the scaling of $\sigma_p$ with $L$ using the powerlaw $\sigma_p = (L/\alpha)^{-\beta} + \sigma^{\infty}_p$. $\sigma^{\infty}_p$ is reported in the main text. (a) $\alpha$ versus $T^*$. System parameters are indicated in the legend. (b) $\beta$ versus $T^*$ for the same systems. Error bars represent the standard error in the fit of $\alpha$ and $\beta$, respectively.}
\end{figure}

